\documentclass[MSNbibl,nameyear,dvips]{arxstspdf}
\usepackage{flushend}
\usepackage{stfloats}
\volume{29}
\issue{1}
\pubyear{2014}
\firstpage{95}
\lastpage{96}
\doi{10.1214/13-STS448} % kopijuoti is 'New paper accepted'
\referstodoi{10.1214/12-STS413}% pagrindinio straipsnio DOI, kai
\makeatletter
\renewcommand{\cite}{\citet}
\makeatother

\begin{document}
\begin{frontmatter}
\vspace*{6pt}
\title{Discussion of ``Estimating the Distribution of Dietary
Consumption Patterns''}%\thanksref{T1}
% kai straipsnis turi susijusiu diskusiju ir rejoinder'iu
%rejoinder at \relateddoi{r}{10.1214/00-STSXXXX}.}
\runtitle{Dietary Consumption Patterns}

\begin{aug}
\author[a]{\fnms{Stephen E.} \snm{Fienberg}\corref{}\ead[label=e1]{fienberg@stat.cmu.edu}}
\and
\author[a]{\fnms{Rebecca C.} \snm{Steorts}\ead[label=e2]{beka@cmu.edu}}
\runauthor{S. E. Fienberg and R. C. Steorts}

\affiliation{Carnegie Mellon University}

\address[a]{Stephen E. Fienberg is Maurice Falk University Professor of Statistics and
Social Science,
Department of Statistics, Machine Learning Department, and
 Heinz College, Carnegie Mellon University, Pittsburgh, Pennsylvania 15213, USA
 \printead{e1}. Rebecca C. Steorts is Visiting Assistant Professor,
 Department of Statistics, Carnegie Mellon University, Pittsburgh, Pennsylvania 15213, USA
\printead{e2}.}

\end{aug}

% ABSTRACT

% KEYWORDS

\end{frontmatter}

Carroll describes an innovative model developed in \cite{zhang_2011}
for estimating dietary consumption patterns in children, and a
successful Bayesian solution for inferring the features of the model.
The original authors went to great lengths to achieve valid frequentist
inference via a Bayesian analysis that simplified the computational
complexities encountered in standard frequentist approaches.
Pragmatically, this led to a reasonable set of estimates, but their
combination of Bayesian and frequentist tools and ideas stopped short
of what we consider a full and proper Bayesian analysis.
%There are good and bad merits to such a strategy.
%One one hand, they achieved a practical answer, however, on the other
%hand,
%in combining Bayes and frequentist tools and ideas, they stopped short
%of what we would consider to be a full and proper Bayesian analysis.
We ask two fundamental questions: How do we know that the model and
estimation are valid? What role should the survey weights have played?

% The good part was they achieved a practical answer. But in combining
%Bayes and frequentist tools and ideas, they stopped short of what we
%would consider to be a full and proper Bayesian analysis. We ask two
%fundamental questions: How do we know that the model and estimation
%are valid? What role should the survey weights have played?

%s1 #&#
\section{Model Validity}

The model of \cite{zhang_2011} is highly com\-plex---how, without
something like sensitivity analysis, are we to know that it is valid?
%Due to the incredible model complexity, more fully explained in
% valid without some sort of sensitivity analysis?
%%should we cite when bernstain-von mises below?
As for inference, the original authors
rely on the well-known (Bernstein--von Mises)
%close relationship between
asymptotic convergence of
Bayesian posterior means and maximum likelihood estimates to \mbox{develop}
standard errors using balanced repeated replication (BRR). We agree
that their sample size is large for many purposes, however, when the
inverse Fisher information is large, convergence can be slow. Moreover,
this standard convergence result is known to slow down as the number of
parameters grows, failing completely for nonparametric models.
Can we rely on Bernstein--von Mises, at these sample sizes, for this
very complex (and only semi-parametric) model?
%Due to these results and the fact that the model is very complex (and
%nonparametric), does Berstein-von Mises hold?
This is not clear to us.

%s2 #&#
\section{Survey Weights}

In Section~3.3, \cite{carroll_2013} notes that the use of survey
weights in Bayesian analyses is controversial, and then
%proceeds to do a weighted analysis nonetheless, taking his weights
%from the National Center of Health Statistics (NCHS).
he proceeds to use them as reported in by the National Center of
Health Statistics (NCHS) nonetheless to do a weighted analysis.
\cite{fienberg_2009} reminds us that in the NCHS survey context,
weights are
not just used to adjust for unequal selection probabilities, but are
the product of at least three factors:
% used to adjust for factors
%other than unequal probabilities of selection, and they are the
%product of at least three
%components:
\begin{eqnarray*}
w_k &=&\frac{1}{\pi_k}\times\mbox{(nonresponse adjustment)}
\\
&&{} \times \mbox{(post-stratification adjustment)}.
\end{eqnarray*}
The first factor is the inverse of the probability of selection, for
example, taking into account stratification and clustering. The second factor
% in the weight
inflates the sample results to adjust for nonresponse, typically by
invoking the assumption that the missing data are missing at random, at
least within
%selected
chosen
strata or post-strata. The third factor
re-weights the population totals to add up to control totals
coming from another source such as a
census.

\cite{gelman_2007} rightly states: ``Survey weighting is a mess,'' and
this is especially so from a Bayesian perspective.
What weights if any should be used in a Bayesian analysis? In a simple
stratification setting, and where we are estimating a mean or a total,
weighting using $1/\pi_k$ has a Bayesian justification. For more
complex situations, such as the one Carroll describes, the role of the
survey weights is unclear. Bayesian benchmarking is a way to deal with
the third component in the weight formula above, but \cite{ghosh_2013}
point out the tricky nature of the choice of both loss function and
benchmarking weights for small area estimation of complex surveys.
%%introduce that fact that the use of loss function weight and survey
%weight is unclear for SAE in Ghosh and Steorts and first paper. add
%this.
In essence, Carroll and his collaborators appear to be creating a
pseudo-likelihood that adjusts individual contributions by the weights
and then they use a survey-weighted MCMC calculation with uncertainty
estimation coming from balanced repeated replication. This seems
unusually strange to us, and decidedly non-Bayesian in character.

%The likelihood given in Appendix A of Zhang et al. (2011) involves
%these weights, but we are told that the estimates of parameters such
%as $\beta$ and $\Sigma_{\epsilon}$ were ``essentially unchanged''
%between weighting and unweighting. Which set of parameter estimates is
%being relied on here is unclear. Survey weights were however
%unquestionably used in going from the estimates of latent variables
%for each individual in the sample to estimates of the distribution of
%latent variables in the general population of children ages 2--8.
%It is unclear to us after the MCMC method is applied why the authors
%use balanced repeated replication (BRR) weights to develop standard
%errors. Is this the best choice of weights? By construction the BRR
%weights are clearly not independent and most likely not exchangeable,
%however could they be tweaked to be exchangeable? Or is there a more

%The authors learn on frequentist origins that don't serve the Bayesian
%paradigm. Why not assess the uncertainty using a posterior mean
%squared error?

Even if this pseudo-likelihood structure is correct, to be \emph{fully
Bayesian}, the weight $w_k$ associated with the $k$th child
\emph{should be} a random variable. The weights should then have a
prior distribution and a likelihood, and be estimated together with the
other unknown parameters. At the very least, ignoring the variability
in the weights will cause the estimate of population distributions to
seem unduly precise.
%Furthermore, pretending
It may be that a proper Bayesian weighting justification of \cite
{carroll_2013} exists, but
simply hoping
that the frequentist approach
to survey weighting
carries over to the Bayesian setting without change seems problematic.\vspace*{6pt}

%?

\section*{Acknowledgment}

Supported in part by NSF Grant SES-1130706 to Carnegie Mellon University.\

%suskaldyti doi

% imsref loaded by audrone.aklyte, 2014-02-17 10:23:21
%

\end{document}